\input harvmac
\input amssym
\input epsf

\def\unit{\relax{\rm 1\kern-.26em I}}
\def\nada{\relax{\rm 0\kern-.30em l}}



\noblackbox
\def\IL{\relax{\rm I\kern-.18em L}}
\def\IH{\relax{\rm I\kern-.18em H}}
\def\IR{\relax{\rm I\kern-.18em R}}
\def\IC{\relax\hbox{$\inbar\kern-.3em{\rm C}$}}
\def\IZ{\relax\ifmmode\mathchoice
{\hbox{\cmss Z\kern-.4em Z}}{\hbox{\cmss Z\kern-.4em Z}}
{\lower.9pt\hbox{\cmsss Z\kern-.4em Z}} {\lower1.2pt\hbox{\cmsss
Z\kern-.4em Z}}\else{\cmss Z\kern-.4em Z}\fi}
\def\CM {{\cal M}}
\def\CN {{\cal N}}

\def\CL {{\cal L}}

\def\CO {{\cal O}}

\def\CA{{\cal A}}

\def\CM {{\cal M}}
\def\CN {{\cal N}}

\def\CO {{\cal O}}

\def\Tr{{\rm Tr}}

\font\manual=manfnt \def\dbend{\lower3.5pt\hbox{\manual\char127}}

\def\IZ{\relax\ifmmode\mathchoice
{\hbox{\cmss Z\kern-.4em Z}}{\hbox{\cmss Z\kern-.4em Z}}
{\lower.9pt\hbox{\cmsss Z\kern-.4em Z}} {\lower1.2pt\hbox{\cmsss
Z\kern-.4em Z}}\else{\cmss Z\kern-.4em Z}\fi}
\def\half {{1\over 2}}

\def\bar{\overline}

\def\rt2{\sqrt{2}}
\def\irt2{{1\over\sqrt{2}}}

\def\hat{\widehat}
\def\slashchar#1{\setbox0=\hbox{$#1$}           
   \dimen0=\wd0                                 
   \setbox1=\hbox{/} \dimen1=\wd1               
   \ifdim\dimen0>\dimen1                        
      \rlap{\hbox to \dimen0{\hfil/\hfil}}      
      #1                                        
   \else                                        
      \rlap{\hbox to \dimen1{\hfil$#1$\hfil}}   
      /                                         
   \fi}

\def\foursqr#1#2{{\vcenter{\vbox{
    \hrule height.#2pt
    \hbox{\vrule width.#2pt height#1pt \kern#1pt
    \vrule width.#2pt}
    \hrule height.#2pt
    \hrule height.#2pt
    \hbox{\vrule width.#2pt height#1pt \kern#1pt
    \vrule width.#2pt}
    \hrule height.#2pt
        \hrule height.#2pt
    \hbox{\vrule width.#2pt height#1pt \kern#1pt
    \vrule width.#2pt}
    \hrule height.#2pt
        \hrule height.#2pt
    \hbox{\vrule width.#2pt height#1pt \kern#1pt
    \vrule width.#2pt}
    \hrule height.#2pt}}}}
\def\psqr#1#2{{\vcenter{\vbox{\hrule height.#2pt
    \hbox{\vrule width.#2pt height#1pt \kern#1pt
    \vrule width.#2pt}
    \hrule height.#2pt \hrule height.#2pt
    \hbox{\vrule width.#2pt height#1pt \kern#1pt
    \vrule width.#2pt}
    \hrule height.#2pt}}}}
\def\sqr#1#2{{\vcenter{\vbox{\hrule height.#2pt
    \hbox{\vrule width.#2pt height#1pt \kern#1pt
    \vrule width.#2pt}
    \hrule height.#2pt}}}}
\def\square{\mathchoice\sqr65\sqr65\sqr{2.1}3\sqr{1.5}3}

\def\figin{\epsfcheck\figin}\def\figins{\epsfcheck\figins}
\def\epsfcheck{\ifx\epsfbox\UnDeFiNeD
\message{(NO epsf.tex, FIGURES WILL BE IGNORED)}
\gdef\figin##1{\vskip2in}\gdef\figins##1{\hskip.5in}
\else\message{(FIGURES WILL BE INCLUDED)}%
\gdef\figin##1{##1}\gdef\figins##1{##1}\fi}
\def\DefWarn#1{}
\def\figinsert{\goodbreak\midinsert}
\def\ifig#1#2#3{\DefWarn#1\xdef#1{fig.~\the\figno}
\writedef{#1\leftbracket fig.\noexpand~\the\figno}%
\figinsert\figin{\centerline{#3}}\medskip\centerline{\vbox{\baselineskip12pt
\advance\hsize by -1truein\noindent\footnotefont{\bf
Fig.~\the\figno:\ } \it#2}}
\bigskip\endinsert\global\advance\figno by1}

\lref\IntriligatorJJ{
  K.~A.~Intriligator, B.~Wecht,
  ``The Exact Superconformal R Symmetry Maximizes a,''
Nucl.\ Phys.\  {\bf B667}, 183-200 (2003). [hep-th/0304128].
}

\lref\SeibergPQ{
  N.~Seiberg,
  ``Electric - Magnetic Duality in Supersymmetric nonAbelian Gauge Theories,''
Nucl.\ Phys.\  {\bf B435}, 129-146 (1995). [hep-th/9411149].
}

\lref\KutasovIY{
  D.~Kutasov, A.~Parnachev, D.~A.~Sahakyan,
 ``Central charges and U(1)(R) symmetries in N=1 superYang-Mills,''
JHEP {\bf 0311}, 013 (2003). [hep-th/0308071].
}

\lref\IntriligatorMI{
  K.~A.~Intriligator, B.~Wecht,
  ``RG fixed points and flows in SQCD with adjoints,''
Nucl.\ Phys.\  {\bf B677}, 223-272 (2004). [hep-th/0309201].
}

\lref\WittenTW{
  E.~Witten,
  ``Global Aspects of Current Algebra,''
Nucl.\ Phys.\  {\bf B223}, 422-432 (1983).
}

\lref\FrishmanDQ{
  Y.~Frishman, A.~Schwimmer, T.~Banks, S.~Yankielowicz,
  ``The Axial Anomaly and the Bound State Spectrum in Confining Theories,''
Nucl.\ Phys.\  {\bf B177}, 157 (1981).
}

\lref\AppelquistHR{
  T.~Appelquist, A.~G.~Cohen, M.~Schmaltz,
  ``A New Constraint on Strongly Coupled Gauge Theories,''
Phys.\ Rev.\  {\bf D60}, 045003 (1999). [arXiv:hep-th/9901109
[hep-th]].
}
\lref\AnselmiAM{
  D.~Anselmi, D.~Z.~Freedman, M.~T.~Grisaru, A.~A.~Johansen,
  ``Nonperturbative Formulas for Central Functions of Supersymmetric Gauge Theories,''
Nucl.\ Phys.\  {\bf B526}, 543-571 (1998). [hep-th/9708042].
}

\lref\AnselmiYS{
  D.~Anselmi, J.~Erlich, D.~Z.~Freedman, A.~A.~Johansen,
  ``Positivity Constraints on Anomalies in Supersymmetric Gauge Theories,''
Phys.\ Rev.\  {\bf D57}, 7570-7588 (1998). [hep-th/9711035].
}

\lref\CardyCWA{
  J.~L.~Cardy,
  ``Is There a c Theorem in Four-Dimensions?,''
Phys.\ Lett.\  {\bf B215}, 749-752 (1988). }

\lref\OsbornTD{
  H.~Osborn,
  ``Derivation of a Four-Dimensional c Theorem,''
Phys.\ Lett.\  {\bf B222}, 97 (1989).
}

\lref\KutasovIY{
  D.~Kutasov, A.~Parnachev, D.~A.~Sahakyan,
  ``Central Charges and U(1)(R) Symmetries in N=1 SuperYang-Mills,''
JHEP {\bf 0311}, 013 (2003). [hep-th/0308071].
}

\lref\IntriligatorMI{
  K.~A.~Intriligator, B.~Wecht,
  ``RG Fixed Points and Flows in SQCD with Adjoints,''
Nucl.\ Phys.\  {\bf B677}, 223-272 (2004). [hep-th/0309201].
}

\lref\JackEB{
  I.~Jack, H.~Osborn,
  ``Analogs for the C Theorem for Four-dimensional Renormalizable Field Theories,''
Nucl.\ Phys.\  {\bf B343}, 647-688 (1990).
}

\lref\ZamolodchikovGT{
  A.~B.~Zamolodchikov,
  ``Irreversibility of the Flux of the Renormalization Group in a 2D Field Theory,''
JETP Lett.\  {\bf 43}, 730-732 (1986). }

\lref\WessYU{
  J.~Wess, B.~Zumino,
  ``Consequences of Anomalous Ward Identities,''
Phys.\ Lett.\  {\bf B37}, 95 (1971). }

\lref\SchwimmerZA{
  A.~Schwimmer, S.~Theisen,
  ``Spontaneous Breaking of Conformal Invariance and Trace Anomaly Matching,''
Nucl.\ Phys.\  {\bf B847}, 590-611 (2011). [arXiv:1011.0696
[hep-th]].
}

\lref\PhamCR{
  T.~N.~Pham, T.~N.~Truong,
  ``Evaluation of the Derivative Quartic Terms of The Meson Chiral Lagrangian from Forward Dispersion Relation,''
Phys.\ Rev.\  {\bf D31}, 3027 (1985).
}

\lref\AdamsSV{
  A.~Adams, N.~Arkani-Hamed, S.~Dubovsky, A.~Nicolis, R.~Rattazzi,
  ``Causality, Analyticity and an IR Obstruction to UV Completion,''
JHEP {\bf 0610}, 014 (2006). [hep-th/0602178].
}

\lref\DineSW{
  M.~Dine, G.~Festuccia, Z.~Komargodski,
  ``A Bound on the Superpotential,''
JHEP {\bf 1003}, 011 (2010). [arXiv:0910.2527 [hep-th]].
}

\lref\CappelliYC{
  A.~Cappelli, D.~Friedan, J.~I.~Latorre,
  ``C Theorem and Spectral Representation,''
Nucl.\ Phys.\  {\bf B352}, 616-670 (1991).
}

\lref\FrishmanDQ{
  Y.~Frishman, A.~Schwimmer, T.~Banks, S.~Yankielowicz,
  ``The Axial Anomaly and the Bound State Spectrum in Confining Theories,''
Nucl.\ Phys.\  {\bf B177}, 157 (1981).
}

\lref\WittenTW{
  E.~Witten,
  ``Global Aspects of Current Algebra,''
Nucl.\ Phys.\  {\bf B223}, 422-432 (1983).
}

\lref\BirrellIX{
  N.~D.~Birrell, P.~C.~W.~Davies,
  ``Quantum Fields in Curved Space,''
Cambridge, Uk: Univ. Pr. ( 1982) 340p.
}

\lref\MyersTJ{
  R.~C.~Myers, A.~Sinha,
 ``Holographic c-Theorems in Arbitrary Dimensions,''
JHEP {\bf 1101}, 125 (2011).
[arXiv:1011.5819 [hep-th]].
}

\lref\DuffWM{
  M.~J.~Duff,
  ``Twenty Years of the Weyl Anomaly,''
Class.\ Quant.\ Grav.\  {\bf 11}, 1387-1404 (1994).
[hep-th/9308075].
}

\lref\BuchbinderJN{
  I.~L.~Buchbinder, S.~M.~Kuzenko, A.~A.~Tseytlin,
 ``On Low-Energy Effective Actions in N=2, N=4 Superconformal Theories in Four-Dimensions,''
Phys.\ Rev.\  {\bf D62}, 045001 (2000).
[hep-th/9911221].
}

\lref\JafferisZI{
  D.~L.~Jafferis, I.~R.~Klebanov, S.~S.~Pufu, B.~R.~Safdi,
  ``Towards the F-Theorem: N=2 Field Theories on the Three-Sphere,''
JHEP {\bf 1106}, 102 (2011).
[arXiv:1103.1181 [hep-th]].
}

\lref\RiegertKT{
  R.~J.~Riegert,
 ``A Nonlocal Action for the Trace Anomaly,''
Phys.\ Lett.\  {\bf B134}, 56-60 (1984).
}

\lref\FradkinTG{
  E.~S.~Fradkin, A.~A.~Tseytlin,
 ``Conformal Anomaly in Weyl Theory and Anomaly Free Superconformal Theories,''
Phys.\ Lett.\  {\bf B134}, 187 (1984).
}

\lref\TS{
  A.~Schwimmer, S.~Theisen,
 in preparation.}

\lref\KulaxiziJT{
  M.~Kulaxizi, A.~Parnachev,
  ``Energy Flux Positivity and Unitarity in CFTs,''
Phys.\ Rev.\ Lett.\  {\bf 106}, 011601 (2011).
[arXiv:1007.0553 [hep-th]].
}

\lref\KlebanovGS{
  I.~R.~Klebanov, S.~S.~Pufu, B.~R.~Safdi,
 ``F-Theorem without Supersymmetry,''
[arXiv:1105.4598 [hep-th]].
}

\lref\HofmanAR{
  D.~M.~Hofman, J.~Maldacena,
  ``Conformal Collider Physics: Energy and Charge Correlations,''
JHEP {\bf 0805}, 012 (2008).
[arXiv:0803.1467 [hep-th]].
}

\lref\FradkinYW{
  E.~S.~Fradkin, G.~A.~Vilkovisky,
  ``Conformal Off Mass Shell Extension and Elimination of Conformal Anomalies in Quantum Gravity,''
Phys.\ Lett.\  {\bf B73}, 209-213 (1978).
}

\lref\DeserYX{
  S.~Deser, A.~Schwimmer,
  ``Geometric Classification of Conformal Anomalies in Arbitrary Dimensions,''
Phys.\ Lett.\  {\bf B309}, 279-284 (1993).
[hep-th/9302047].
}

\lref\BonoraCQ{
  L.~Bonora, P.~Pasti, M.~Bregola,
  ``Weyl Cocycles,''
Class.\ Quant.\ Grav.\  {\bf 3}, 635 (1986).
}

\lref\NirSV{
  Y.~Nir,
 ``Infrared Treatment of Higher Anomalies and Their Consequences,''
Phys.\ Rev.\  {\bf D34}, 1164-1168 (1986).
}

\lref\DistlerIF{
  J.~Distler, B.~Grinstein, R.~A.~Porto, I.~Z.~Rothstein,
  ``Falsifying Models of New Physics via WW Scattering,''
Phys.\ Rev.\ Lett.\  {\bf 98}, 041601 (2007). [hep-ph/0604255].
}

\lref\KomargodskiVJ{
  Z.~Komargodski, A.~Schwimmer,
  ``On Renormalization Group Flows in Four Dimensions,''
[arXiv:1107.3987 [hep-th]].
}

\lref\FreedmanGP{
  D.~Z.~Freedman, S.~S.~Gubser, K.~Pilch, N.~P.~Warner,
  ``Renormalization group flows from holography supersymmetry and a c theorem,''
Adv.\ Theor.\ Math.\ Phys.\  {\bf 3}, 363-417 (1999).
[hep-th/9904017].
}

\lref\GirardelloBD{
  L.~Girardello, M.~Petrini, M.~Porrati, A.~Zaffaroni,
  ``The Supergravity dual of N=1 superYang-Mills theory,''
Nucl.\ Phys.\  {\bf B569}, 451-469 (2000). [hep-th/9909047].
}

\lref\BanksNN{
  T.~Banks and A.~Zaks,
  ``On The Phase Structure Of Vector-Like Gauge Theories With Massless
  Fermions,''
  Nucl.\ Phys.\  B {\bf 196}, 189 (1982).
}

\lref\JackEB{
  I.~Jack and H.~Osborn,
  ``Analogs for the c Theorem for Four-Dimensional Renormalizable Field
 Theories,''
  Nucl.\ Phys.\  B {\bf 343}, 647 (1990).
}

\lref\GasserGG{
  J.~Gasser and H.~Leutwyler,
  ``Chiral Perturbation Theory: Expansions In The Mass Of The Strange Quark,''
  Nucl.\ Phys.\  B {\bf 250}, 465 (1985).
}

\lref\GasserYG{
  J.~Gasser and H.~Leutwyler,
  ``Chiral Perturbation Theory To One Loop,''
  Annals Phys.\  {\bf 158}, 142 (1984).
}

\lref\NovikovUC{
  V.~A.~Novikov, M.~A.~Shifman, A.~I.~Vainshtein and V.~I.~Zakharov,
  ``Exact Gell-Mann-Low Function Of Supersymmetric Yang-Mills Theories From
  Instanton Calculus,''
  Nucl.\ Phys.\  B {\bf 229}, 381 (1983).
}

\lref\SeibergVC{
  N.~Seiberg,
  ``Naturalness Versus Supersymmetric Non-renormalization Theorems,''
  Phys.\ Lett.\  B {\bf 318}, 469 (1993)
  [arXiv:hep-ph/9309335].
}

\lref\AntoniadisGN{
  I.~Antoniadis and M.~Buican,
  ``On R-symmetric Fixed Points and Superconformality,''
  Phys.\ Rev.\  D {\bf 83}, 105011 (2011)
  [arXiv:1102.2294 [hep-th]].
}

\lref\DorigoniRA{
  D.~Dorigoni and V.~S.~Rychkov,
 ``Scale Invariance + Unitarity $\rightarrow$ Conformal Invariance?,''
  arXiv:0910.1087 [hep-th].
}

\lref\FortinSZ{
  J.~F.~Fortin, B.~Grinstein, A.~Stergiou,
  ``Scale without Conformal Invariance: Theoretical Foundations,''
[arXiv:1107.3840 [hep-th]].
}

\lref\NakayamaFE{
  Y.~Nakayama,
  ``No Forbidden Landscape in String/M-theory,''
JHEP {\bf 1001}, 030 (2010).
[arXiv:0909.4297 [hep-th]].
}

\lref\GreenDA{
  D.~Green, Z.~Komargodski, N.~Seiberg, Y.~Tachikawa and B.~Wecht,
 ``Exactly Marginal Deformations and Global Symmetries,''
  JHEP {\bf 1006}, 106 (2010)
  [arXiv:1005.3546 [hep-th]].
}

\lref\NakayamaWQ{
  Y.~Nakayama,
  ``On $\epsilon$-conjecture in a-theorem,''
[arXiv:1110.2586 [hep-th]].
}

\lref\progress{
  work in progress }

\lref\PolchinskiDY{
  J.~Polchinski,
  ``Scale and Conformal Invariance in Quantum Field Theory,''
  Nucl.\ Phys.\  B {\bf 303}, 226 (1988).
}

\lref\SchwimmerZA{
  A.~Schwimmer and S.~Theisen,
  ``Spontaneous Breaking of Conformal Invariance and Trace Anomaly Matching,''
  Nucl.\ Phys.\  B {\bf 847}, 590 (2011)
  [arXiv:1011.0696 [hep-th]].
}

\lref\CurtrightQG{
  T.~L.~Curtright, X.~Jin and C.~K.~Zachos,
  ``RG flows, cycles, and c-theorem folklore,''
[arXiv:1111.2649 [hep-th]].
}

\lref\BahJE{
  I.~Bah and B.~Wecht,
  ``New N=1 Superconformal Field Theories In Four Dimensions,''
  arXiv:1111.3402 [hep-th].
}

\lref\IntriligatorJJ{
  K.~A.~Intriligator and B.~Wecht,
 ``The exact superconformal R-symmetry maximizes a,''
  Nucl.\ Phys.\  B {\bf 667}, 183 (2003)
  [arXiv:hep-th/0304128].
}

\lref\MyersTJ{
  R.~C.~Myers and A.~Sinha,
 ``Holographic c-theorems in arbitrary dimensions,''
  JHEP {\bf 1101}, 125 (2011)
  [arXiv:1011.5819 [hep-th]].
}

\lref\CasiniBW{
  H.~Casini and M.~Huerta,
 ``A finite entanglement entropy and the c-theorem,''
  Phys.\ Lett.\  B {\bf 600}, 142 (2004)
  [arXiv:hep-th/0405111].
}

\lref\JafferisZI{
  D.~L.~Jafferis, I.~R.~Klebanov, S.~S.~Pufu and B.~R.~Safdi,
  ``Towards the F-Theorem: N=2 Field Theories on the Three-Sphere,''
  JHEP {\bf 1106}, 102 (2011)
  [arXiv:1103.1181 [hep-th]].
}

\lref\CasiniKV{
  H.~Casini, M.~Huerta and R.~C.~Myers,
  ``Towards a derivation of holographic entanglement entropy,''
JHEP\ {\bf 1105}, 036  (2011).
[arXiv:1102.0440 [hep-th]].
}

\lref\SolodukhinDH{
  S.~N.~Solodukhin,
  ``Entanglement entropy, conformal invariance and extrinsic geometry,''
  Phys.\ Lett.\  B {\bf 665}, 305 (2008)
  [arXiv:0802.3117 [hep-th]].
}

\lref\DuffWM{
  M.~J.~Duff,
  ``Twenty years of the Weyl anomaly,''
  Class.\ Quant.\ Grav.\  {\bf 11}, 1387 (1994)
  [arXiv:hep-th/9308075].
}

\lref\GreenDA{
  D.~Green, Z.~Komargodski, N.~Seiberg, Y.~Tachikawa and B.~Wecht,
  ``Exactly Marginal Deformations and Global Symmetries,''
  JHEP {\bf 1006}, 106 (2010)
  [arXiv:1005.3546 [hep-th]].
}

\lref\AlvarezGaumeIG{
  L.~Alvarez-Gaume and E.~Witten,
  ``Gravitational Anomalies,''
Nucl.\ Phys.\ B\ {\bf 234}, 269  (1984)..
}

\lref\DeserYX{
  S.~Deser and A.~Schwimmer,
  ``Geometric classification of conformal anomalies in arbitrary dimensions,''
  Phys.\ Lett.\  B {\bf 309}, 279 (1993)
  [arXiv:hep-th/9302047].
}

\lref\BuicanTY{
  M.~Buican,
  ``A Conjectured Bound on Accidental Symmetries,''
  arXiv:1109.3279 [hep-th].
}

\lref\PhamCR{
  T.~N.~Pham, T.~N.~Truong,
  ``Evaluation of the Derivative Quartic Terms of The Meson Chiral Lagrangian from Forward Dispersion Relation,''
Phys.\ Rev.\  {\bf D31}, 3027 (1985).
}

\lref\AdamsSV{
  A.~Adams, N.~Arkani-Hamed, S.~Dubovsky, A.~Nicolis, R.~Rattazzi,
  ``Causality, Analyticity and an IR Obstruction to UV Completion,''
JHEP {\bf 0610}, 014 (2006). [hep-th/0602178].
}

\lref\DineSW{
  M.~Dine, G.~Festuccia, Z.~Komargodski,
  ``A Bound on the Superpotential,''
JHEP {\bf 1003}, 011 (2010). [arXiv:0910.2527 [hep-th]].
}

\lref\DistlerIF{
  J.~Distler, B.~Grinstein, R.~A.~Porto, I.~Z.~Rothstein,
  ``Falsifying Models of New Physics via WW Scattering,''
Phys.\ Rev.\ Lett.\  {\bf 98}, 041601 (2007). [hep-ph/0604255].
}

\lref\AdamsHP{
  A.~Adams, A.~Jenkins and D.~O'Connell,
  ``Signs of analyticity in fermion scattering,''
  arXiv:0802.4081 [hep-ph].
}


\rightline{WIS/12/11-DEC-DPPA} \Title{
} {\vbox{\centerline{ The Constraints of Conformal Symmetry on RG Flows  }}}
\medskip

\centerline{\it Zohar Komargodski }
\bigskip
\centerline{ Weizmann Institute of Science, Rehovot
76100, Israel}
 \centerline{
Institute for Advanced Study, Princeton, NJ 08540, USA}

\smallskip

\vglue .3cm
\bigskip
\bigskip
\bigskip
\noindent 

If the coupling constants in QFT are promoted to functions of space-time, the dependence of the path integral on these couplings  is highly constrained by conformal symmetry. We begin the present note by showing that this idea leads to a new proof of Zamolodchikov's theorem. We then review how this simple observation also leads to a derivation of the $a$-theorem. We exemplify the general procedure in some interacting theories in four space-time dimensions. We concentrate on Banks-Zaks and weakly relevant flows, which can be controlled by ordinary and conformal perturbation theories, respectively.  We compute explicitly the dependence of the path integral on the coupling constants and extract the change in the $a$-anomaly (this agrees with more conventional computations of the same quantity). We also discuss some general properties of the sum rule found in~\KomargodskiVJ\ and study it in several examples.

\Date{Dec 2011}

\newsec{Introduction, a New Proof of the c-Theorem, and Summary}

Promoting various coupling constants to background fields has always been a useful tool in the analysis of QFT. The applications of this idea are too numerous to list exhaustively. For example, one may recall the classification of terms in the pion Lagrangian by  Gasser and Leutwyler~\refs{\GasserYG,\GasserGG}. In addition, many of the seminal realizations regarding the dynamics of SUSY gauge theories can also be understood by studying the dependence of various observables on coupling constants. The NSVZ beta function~\NovikovUC\ is one such example, and Seiberg's realization  regarding the power of holomorphy~\SeibergVC\ is another.

Let us recall why this is such a powerful idea. Generically, when various parameters in the Lagrangian are set to zero one finds that the symmetries of the theory are enhanced. If one reintroduces the coupling constants, this enhanced symmetry breaks explicitly to the actual symmetry of the theory. However, we can always assign transformation rules to the coupling constants such that, if the field transformations are accompanied by transformations of the coupling constants, the full {\it enhanced} symmetry is preserved. One  can easily see that, for instance, the expectation values of operators have to be consistent with the enhanced symmetry. In addition, when one integrates over the fluctuating fields, one obtains a functional of the background parameters. This functional of the background parameter ought to respect the full extended symmetry. 

The statements above can be slightly generalized. When the coupling constants are set to zero, some of the extended symmetries of the Lagrangian could have quantum anomalies. Then, under a transformation of the fields by this extended symmetry one picks up the anomaly. Introducing the couplings back to the Lagrangian and letting them transform under the the extended symmetries, one does not create new sources for the violation of the extended symmetry. In other words, performing the extended symmetry transformation on the theory, one still picks up exclusively the anomaly. Then, for instance, if we path integrate over the dynamical fields and remain with a functional of the background parameters, it must be true that this functional of the background parameters reproduces the anomaly.

Imagine any renormalizable QFT (in any number of dimensions) and set all the mass parameters to zero. The extended symmetry includes the full conformal group.\foot{Our discussion hereafter applies to scale invariant theories which are also conformal. It is not known whether all scale invariant theories are conformal (under some assumptions the answer in two dimensions is positive~\PolchinskiDY) and it is not clear which of our results henceforth, if any, would apply to a scale invariant theory which is not conformal (if such an example existed). The reader interested in this topic is referred to some recent literature on related matters~\refs{\NakayamaFE,\DorigoniRA,\AntoniadisGN,\FortinSZ,\CurtrightQG}. } If the number of space-time dimensions is even then the conformal group has trace anomalies. (For a review of trace anomalies see~\DuffWM, and for a classification of trace anomalies see~\DeserYX.) In addition, if the number of space-time dimensions is of the fom $4k+2$, there may be gravitational anomalies~\AlvarezGaumeIG. We will completely ignore gravitational anomalies here. 

Upon introducing the mass terms, one violates conformal symmetry {\it explicitly}. Thus, in general, the conformal symmetry is violated both by trace anomalies and by an operatorial violation of the equation $T_\mu^\mu=0$ in flat space-time. As we explained, on general grounds, the latter violation can always be removed by letting the coupling constants transform. 
Indeed, replace every mass scale $M$ (either in the Lagrangian or associated to some cutoff)  by $Me^{-\tau(x)}$, where $\tau(x)$ is some background field (i.e.~a function of space-time). Then the conformal symmetry of the Lagrangian is restored if we accompany the ordinary conformal transformation of the fields  by a transformation of $\tau$.  To linear order, $\tau(x)$ always appears in the Lagrangian as $\sim \int d^dx \ \tau T_\mu^\mu$. Setting $\tau=0$ one is back to the original theory, but we can also let $\tau$ be some general function of space-time. The variation of the path integral under such a conformal transformation that also acts on $\tau(x)$ is thus fixed by the anomaly of the conformal theory in the ultraviolet. This procedure allows us to study some questions about general RG flows using the constraints of conformal symmetry.

One such question is the dependence on $\tau$ at very low energies. In other words, we integrate out all the high energy modes and flow to the deep infrared. Since we do not integrate out the massless particles, the dependence on $\tau$ is regular and local. As we have explained, the dependence on $\tau$ is tightly constrained by the conformal symmetry. Since in even dimensions the conformal group has trace anomalies, these must be reproduced by the low energy theory. The conformal field theory at long distances, CFT$_{IR}$, contributes to the trace anomalies, but to match to the defining theory, the dilaton functional has to compensate precisely for the difference between the anomalies of the conformal field theory at short distances, CFT$_{UV}$, and the conformal field theory at long distances, CFT$_{IR}$. 

To warm up, let us see how these general ideas are borne out in two-dimensional renormalization group flows. First, let us study the constraints imposed by conformal symmetry on the action of $\tau$ (remember $\tau$ is a background field everywhere in this paper). An easy way to analyze these constraints is to introduce a fiducial metric $g_{\mu\nu}$ into the system. Weyl transformations act on the dilaton and metric according to $\tau\rightarrow \tau+\sigma$, $g_{\mu\nu}\rightarrow e^{2\sigma}g_{\mu\nu}$. If the Lagrangian for the dilaton and metric is Weyl invariant, upon setting the metric to be flat, one finds a conformal invariant theory for the dilaton. Hence, the task is to classify local diff$\times$ Weyl invariant Lagraignains for the dilaton and metric background fields. 

It is convenient to define $\hat g_{\mu\nu}=e^{-2\tau}g_{\mu\nu}$, which is Weyl invariant. At the level of two derivatives, there is only one diff$\times$Weyl invariant term: $\int \sqrt {\hat g} \hat R$. However, this is a topological term, and so it is insensitive to local changes of $\tau(x)$. Therefore, if one starts from a diff$\times$Weyl invariant theory, upon setting $g_{\mu\nu}=\eta_{\mu\nu}$, the term $\int d^2x (\del\tau)^2$ is absent.
 
The key is to recall that unitary two-dimensional theories have a  trace anomaly
\eqn\tractano{T_\mu^\mu=-{c\over 24\pi}R~.}
(In this convention a free scalar field has $c=1$.)
One must therefore allow the Lagrangian to break Weyl invariance, such that the Weyl variation of the action is consistent with~\tractano.
The  action functional which reproduces the two-dimensional trace anomaly is
\eqn\twod{S_{WZ}[\tau,g_{\mu\nu}]= {c\over 24\pi}\int\sqrt g\left(\tau R+(\del\tau)^2\right)~.}
We see that even though the anomaly itself disappears in flat space~\tractano, there is a two-derivative term for $\tau$ that survives even after the metric is taken to be flat. This is of course the familiar Wess-Zumino term for the two-dimensional conformal group.

Consider now some general two-dimensional RG flow from a CFT in the UV (with central charge $c_{UV}$ and a CFT in the IR (with central change $c_{IR}$). Replace every mass scale according to $M\rightarrow Me^{-\tau(x)}$.  We also couple the theory to some background metric.
Performing a simultaneous Weyl transformation of the dynamical fields and the background field $\tau(x)$, the theory is non-invariant only because of the anomaly $\delta_\sigma S={c_{UV}\over 24}\int d^2x\sqrt g\sigma R$. Since this is a property of the full quantum theory, it must be reproduced at all scales.  An immediate consequence of this idea is that also in the deep infrared the effective action should reproduce the transformation $\delta_\sigma S={c_{UV}\over 24}\int d^2x\sqrt g\sigma R$. At long distances, one obtains a contribution $c_{IR}$ to the anomaly from CFT$_{IR}$, hence, the rest of the anomaly must come from  an explicit Wess-Zumino functional~\twod\ with coefficient  $c_{UV}-c_{IR}$.  In particular, setting the background metric to be flat, we conclude that the low energy theory must contain a term 
\eqn\termlow{{c_{UV}-c_{IR}\over 24\pi}\int d^2x (\del\tau)^2~.}
Note that the coefficient of this term is universally proportional to the difference between the anomalies and it does not depend on the details of the flow. Higher-derivative terms for the dilaton can be generated from local diff$\times$Weyl invariant terms, and  there is no a priori reason for them to be universal  (that is, they may depend on the details of the flow, and not just on the conformal field theories at short and long distances).

Zamolodchikov's theorem~\ZamolodchikovGT\ follows directly from~\termlow.  Indeed, we consider the partition function of the ({\it Euclidean}) theory in the presence of two insertions of the background $\tau(x)$, as in figure 1. 

\medskip
\epsfxsize=2.5in \centerline{\epsfbox{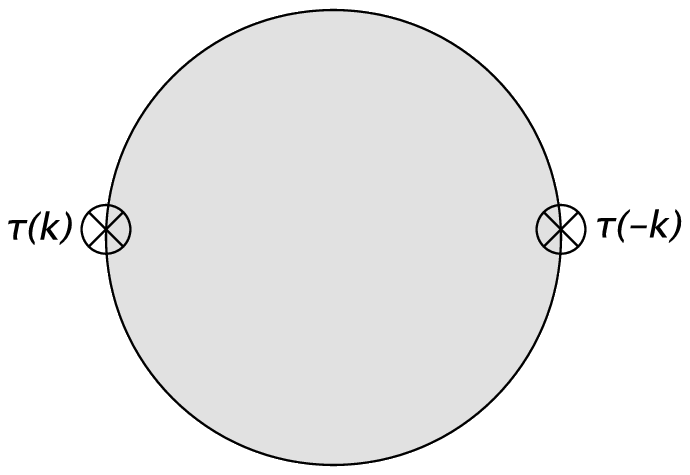}}
\noindent
Fig.1: The partition function of the Euclidean theory with two insertions of the background field with  momentum $k$.
\medskip

From this general object we can extract $c_{UV}-c_{IR}$ by expanding around $k=0$, reading out the term quadratic in momentum, and matching to~\termlow. Reflection positivity thus immediately leads to
\eqn\thethm{c_{UV}>c_{IR}~.} 

We can be more explicit about what precisely goes into the calculation of figure 1.
The coupling of $\tau$ to matter must take the form $\tau T_{\mu}^\mu+\cdots$, where the corrections have more $\tau$s. To extract the two-point function of $\tau$ with two derivatives we must use the insertion $\tau T_\mu^\mu$ twice. (Terms containing $\tau^2$ can be lowered once, but they do not contribute to the two-derivative term in the effective action of $\tau$.)  As a consequence, we find that 
\eqn\cpt{\eqalign{&\bigl\langle e^{\int\tau T_\mu^\mu d^2x}\bigr\rangle =\cdots+\half\int\int \tau(x)\tau(y) \langle T_\mu^\mu(x) T_\mu^\mu(y)\rangle d^2xd^2y+\cdots\cr&=
\cdots+{1\over 4}\int \tau(x)\del_\rho\del_\sigma\tau(x)\left(\int (y-x)^\rho(y-x)^\sigma \langle T_\mu^\mu(x) T_\mu^\mu(y)\rangle d^2y\right)d^2x+\cdots~.}}
In the final line of the equation above, we have concentrated entirely on the two-derivative term.
It follows from translation invariance that the $y$ integral is $x$-independent  
\eqn\easyeq{\int (y-x)^\rho(y-x)^\sigma \langle T_\mu^\mu(x) T_\mu^\mu(y)\rangle d^2y=\half\eta^{\rho\sigma}\int y^2 \langle T_\mu^\mu(0) T_\mu^\mu(y)\rangle  d^2y~.}

To summarize, one finds the following contribution to the dilaton effective action at two derivatives
\eqn\dilefftwod{{1\over 8}\int d^2x\tau\square  \tau\int d^2yy^2\langle T(y)T(0)\rangle~.}
According to~\termlow, the expected coefficient of $\tau\square\tau$ is $(c_{UV}-c_{IR})/24\pi$, and so by comparing we obtain 
\eqn\final{\Delta c=3\pi\int d^2yy^2\langle T(y)T(0)\rangle~.}
As we have already mentioned, $\Delta c>0$ follows from reflection positivity (which is a property of unitary theories). Equation~\final\ agrees with the classic results about two-dimensional flows, for instance the sum rule in~\CappelliYC. (One has to remember that their definition of the stress tensor differs by a factor of $2\pi$ from ours.)  

This finishes our discussion of two-dimensional flows. The same ideas can be applied to four-dimensional flows. The following presentation of the proof of the $a$-theorem is completely equivalent to the one in~\KomargodskiVJ. The only slight pedagogical difference being that we treat $\tau$ as a {\it background c-number field} and never path integrate over it. In this way one avoids having to introduce the, artificial, large decay constant, and therefore one does not need to expand in $1/f$. This difference is purely pedagogical as far as the dynamics of QFT in 3+1 dimensions is concerned, but this slight twist in the logic allows to exhibit the harmony between two and four dimensions. 

One starts by classifying local diff$\times$Weyl invariant functionals of $\tau$ and a background metric $g_{\mu\nu}$. It has been shown that the result up to four derivatives is (setting the background metric to be flat)
\eqn\geneffaction{\eqalign{\int d^4x\left(\alpha_1 e^{-4\tau}+\alpha_2(\del e^{-\tau})^2+\alpha_3\left(\square \tau-(\del\tau)^2\right)^2\right)~,}}
where $\alpha_i$ are some real coefficients. However, in the quantum theory Weyl invariance must be violated because of the $a$- and $c$-trace anomalies 
\eqn\eqntraceanomalies{T_\mu^\mu=aE_4-cW_{\mu\nu\rho\sigma}^2~,}  
where $E_4$ is the Euler density and $W_{\mu\nu\rho\sigma}$ is the Weyl tensor. It turns out that the $c$-anomaly does not lead to a Wess-Zumino term, so the $c$-anomaly disappears when the background metric is flat. The $a$-anomaly, however, does lead to a Wess-Zumino term (see~\SchwimmerZA,\KomargodskiVJ). Since the total anomaly in the infrared must match that of the defining theory, the Wess-Zumino term again comes with a universal coefficient
\eqn\WZfourd{S_{WZ}= 2(a_{UV}-a_{IR})\int d^4x\left(2(\del\tau)^2\square\tau-(\del\tau)^4\right)~.}

We see that if one knew the four-derivative terms for the dilaton, by comparing~\geneffaction\ and \WZfourd, one could extract $a_{UV}-a_{IR}$.  A more transparent way to discern the WZ term from the term proportional to $\alpha_3$ in~\geneffaction\ is found by switching to a new variable 
\eqn\newvar{\Psi=1-e^{-\tau}~.} 
In terms of this variable~\geneffaction\ becomes
\eqn\fourdernewvar{\int d^4x\left(\alpha_1\Psi^4+\alpha_2(\del\Psi)^2+{\alpha_3\over (1-\Psi)^2}(\square\Psi)^2\right)~,}
while the WZ term~\WZfourd\ is 
\eqn\WZfourdnewvar{S_{WZ}=2(a_{UV}-a_{IR})\int d^4x\left({2(\del\Psi)^2\square\Psi\over (1-\Psi)^3}+{(\del\Psi)^4\over (1-\Psi)^4}\right)~.}
The presence of $\Psi$ in the denominator is not alarming since~\fourdernewvar,\WZfourdnewvar\ are to be understood as an expansion in derivatives and in $\Psi$, around $\Psi=0$.

We see that if we consider background fields $\Psi$ which are null ($\square\Psi=0$) $\alpha_3$ disappears and only the last term in~\WZfourdnewvar\ remains. Therefore, by computing the partition function of the QFT in the presence of four null insertions of $\Psi$ one can extract directly $a_{UV}-a_{IR}$.

\medskip
\epsfxsize=2.5in \centerline{\epsfbox{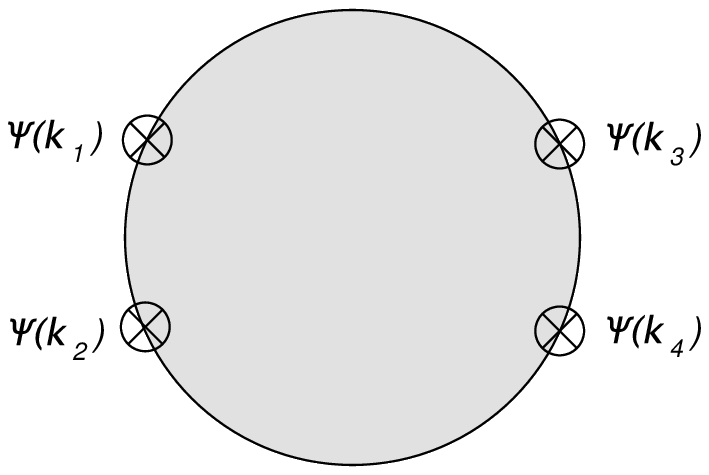}}
\noindent
Fig.2: Four insertions of the background field $\Psi$ with $\sum_ik_i=0$ and $k_i^2=0$. The blob represents the quantum matter fields.
\medskip

Indeed, consider all the diagrams with four insertions of a background $\Psi$ with momenta $k_i$, such that $\sum_ik_i=0$ and $k_i^2=0$ (see figure 2). 
Expanding this amplitude to fourth order in the momenta $k_i$, one finds that the momentum dependence takes the form $s^2+t^2+u^2$ with $s=2k_1\cdot k_2$, $t=2k_1\cdot k_3$, $u=2k_1\cdot k_4$. Our effective action analysis shows that the coefficient of $s^2+t^2+u^2$ is directly proportional to $a_{UV}-a_{IR}$.

In fact, one can even specialize to the so-called forward kinematics, choosing $k_1=-k_3$ and $k_2=-k_4$. Then the amplitude of figure 2 is only a function of $s=2k_1\cdot k_2$.  $a_{UV}-a_{IR}$ can be extracted from the $s^2$ term in the expansion of the amplitude around $s=0$. Continuing $s$ to the complex plane, there is a branch cut for positive $s$ (corresponding to physical states in the $s$-channel) and negative $s$ (corresponding to physical states in the $u$-channel). There is a crossing symmetry $s\leftrightarrow -s$ so these branch cuts are identical.

To calculate the imaginary part associated to the branch cut we utilize the optical theorem. See figure 3. The imaginary part is manifestly positive definite.

\medskip
\epsfxsize=4.7in \centerline{\epsfbox{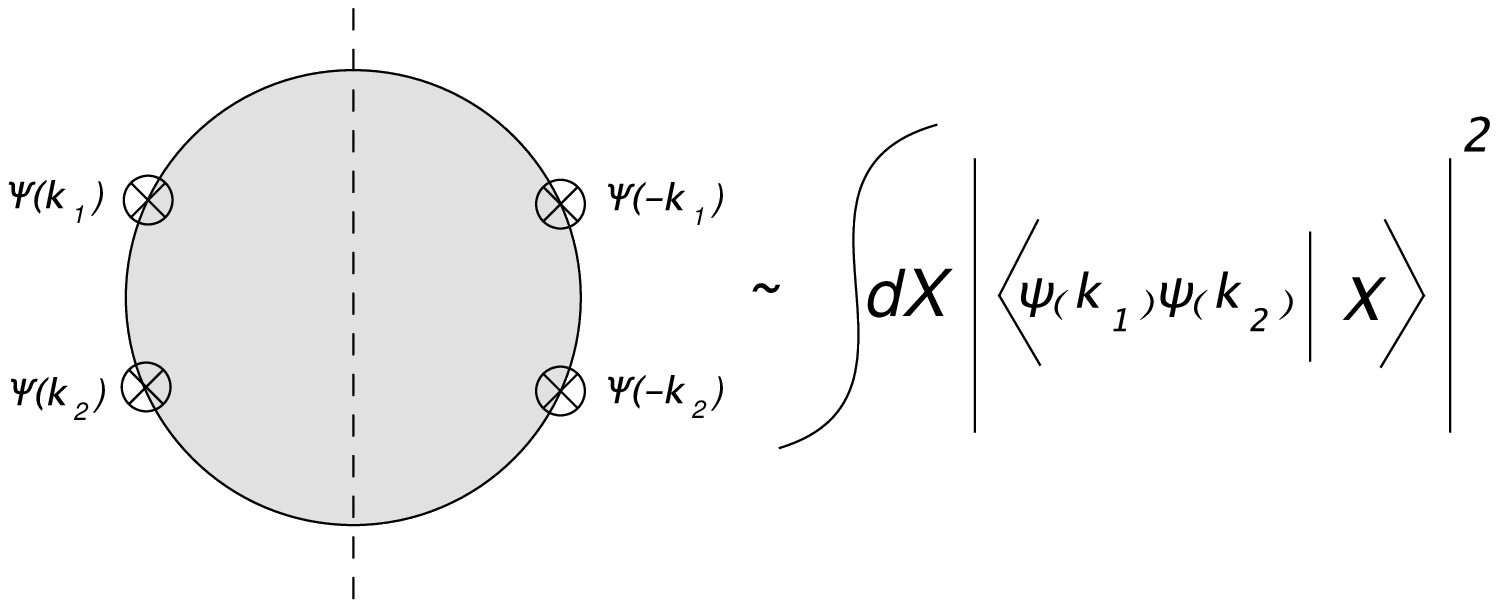}}
\noindent
Fig.3: The imaginary part is given by calculating all the connected diagrams involving two insertions of the background field and any final state. One then squares the amplitude for the transition to this particular final state and sums over all possible final states.
\medskip

Using Cauchy's theorem we can relate the low energy coefficient of $s^2$, $a_{UV}-a_{IR}$, to an integral over the branch cut. Fixing all the coefficients one finds \eqn\sumrule{a_{UV}-a_{IR}={1\over 4\pi}\int_{s>0}{Im\CA(s) \over s^3 }~.}
As explained, the imaginary part $Im\CA(s)$ can be evaluated by means of the optical theorem, figure 3, and hence it is manifestly positive. Since the integral converges by power counting (and thus no subtractions are needed), we conclude 
\eqn\cardyconj{a_{UV}>a_{IR}~.} 
Analyticity arguments of this type have appeared in other applications before.\foot{For instance, in the context of the pion Lagrangian see~\PhamCR, and in the closely related realm of electroweak physics the analysis is carried out in~\DistlerIF. There are also applications for supersymmetric theories~\DineSW, and for fermion scattering~\AdamsHP. A refreshing point of view 
on the nature of these constraints was given in~\AdamsSV.
}

It is interesting to compare the proof of the $a$-theorem and the argument regarding two-dimensional flows. As we have emphasized repeatedly, they both rely on the same simple idea of promoting the masses to functions of space-time.  In addition, in both arguments the key is to identity the anomalous Wess-Zumino-like term in the generating functional for the background dilaton. This allows us to isolate a special term in the dilaton functional which only depends on the anomalies in the UV and IR CFTs, and not on the particular flow. The main difference is, however, in the way positivity is established. In two dimensions, one invokes reflection positivity of a two-point function 
(reflection positivity is best understood in {\it Euclidean} space). In four dimensions, the Wess-Zumino term 
involves four dilatons, so the natural positivity constraint comes from the forward kinematics (and hence, it is inherently {\it Minkowskian}).

Let us say a few words about the physical relevance of~\cardyconj. Such an inequality constrains severely the dynamics of quantum field theory, and in favorable cases can be used to establish that some symmetries must be broken or that some symmetries must be unbroken. In a similar fashion, if a system naively admits several possible dynamical scenarios one can use~\cardyconj\ as an additional handle. (For example, an interplay between $a$-maximization~\IntriligatorJJ\ and the $a$-theorem has been used in~\BahJE\ to throw light on the possible dynamics of various $\CN=1$ theories.\foot{An additional constraint on a class of supersymmetric theories has been proposed recently in~\BuicanTY. It would be very nice to understand it in the language of the effective action for background fields.})  The $a$-anomaly is also closely related to the entanglement entropy across spheres~\refs{\SolodukhinDH,\MyersTJ,\CasiniKV} and it would be very interesting to understand the inequality~\cardyconj\ in these terms. (In two dimensions it has already been shown~\CasiniBW\  that one can derive Zamolodchikov's theorem via the ordinary inequalities that entanglement entropy obeys.) A concrete relation to the entanglement entropy is also likely to be generalizable to 2+1 dimensions, where a plausible suggestion of a universal monotonic property of renormalization group flows now exists~\refs{\MyersTJ,\JafferisZI}.

Note that even though we have already shown that the $a$-theorem holds, and further, one can easily construct a universal monotonically decreasing function that interpolates between $a_{UV}$ and $a_{IR}$ (e.g. by cutting the integral~\sumrule\ at intermediate scales; obviously there are infinitely many ways of doing this), one fundamental question about four-dimensional  renormalization group flows is still outstanding. That is, whether one can say something akin to the gradient flow property in two dimensions. The property of gradient flow in two dimensions follows directly from the fact that there is a two-point function in the sum rule~\final. The situation in four dimensions is more complicated; one deals with a four-point function which lives naturally in Minkowski space. We hope to elucidate the object that appears instead of the familiar gradient flow soon. 

In the remaining of this note  we consider some interacting models in four space-time dimensions. We being from generic weakly relevant flows and continue to the Banks-Zaks~\BanksNN\ fixed point. In both cases we perform the path integral 
over the quantum fields and extract the dependence on the coupling constants. Consistency requires the dependence on the coupling constants assumes the constrained form~\geneffaction,\WZfourd. We verify that this is indeed the case and calculate $a_{UV}-a_{IR}$ in these examples. In both cases our result for $a_{UV}-a_{IR}$ agrees with other, more conventional, methods of extracting the change in the $a$-anomaly.  These computations are summarized in sections~2,3.  

In section~4 we discuss in more details the sum rule~\sumrule.  We first explain how the integrand $Im\CA(s)/s^3$ behaves at small and large $s$ in a general quantum field theory. We then examine the sum rule explicitly in the Banks-Zaks flow (and also revisit the  free massive scalar).

\newsec{Weakly Relevant Flows}

Consider a four-dimensional CFT in which there exists a primary operator $\CO$ of dimension $\Delta=4-\epsilon$, with $\epsilon>0$. Let us deform the action by this operator,   $\delta S=\int d^4x \lambda \CO(x)$. The beta function for the dimensionless coupling $g=\lambda\mu^{-\epsilon}$ takes the form
 \eqn\betafun{{dg\over d\log\mu}=-g\epsilon +\half C\Omega_3g^2+\cdots~,}
where the~$\cdots$ stand for corrections of higher order in $g$, $\Omega_{d-1}\equiv 2\pi^{d/2}/\Gamma(d/2)$, and  $C$ is the coefficient in the OPE
\eqn\OPE{\CO(x)\CO(y)\sim {1\over (x-y)^{8-2\epsilon}  }+C{\CO(x)\over (x-y)^{4-\epsilon}}+\cdots~.}

If one assumes that $\epsilon<<1$ and that $C>0$, there exists an IR fixed point with $g_*=2\epsilon/C\Omega_3$. The smallness of $g_*$ guarantees that this fixed point can be controlled by conformal perturbation theory around $g=0$ and that we can henceforth drop the corrections in~\betafun.

\subsec{Promoting Coupling Constants to Functions of Space-Time}

We restore conformal invariance by promoting the coupling constant $g$ to a function of space-time. A convenient way to do this is to replace the perturbation $g(\mu)\mu^{\epsilon}\CO$ by 
\eqn\pertCFT{\delta S=\int d^4x g\left(F(x)\mu\right)\mu^\epsilon \CO(x)~,}
where 
\eqn\omegadil{F= e^\tau~.}
This theory is now formally conformal because $e^\tau$ transforms like a coordinate under scale transformations. In other words, the renormalization group transformation taking the theory from the scale $\mu$ to $\mu'$ can be compensated by the shift of the dilaton, such that  the physical coupling remains intact. 

Note that to linear order in the dilaton,~\pertCFT\ means that we have $\sim \tau \beta(g)\mu^\epsilon\CO$, which is the expected coupling to the trace of the energy-momentum tensor.
The prescription~\pertCFT,\omegadil\ is  valid under any circumstances (the generalization to CFTs deformed by several relevant operators is obvious),  but we can compute the path integral explicitly and extract the dependence on $\tau$ only in special cases. 

In the weakly relevant flows we are discussing here the coupling $g$ stays small throughout the RG evolution and so we can expand the (Euclidean) partition function as follows \eqn\CFTdil{\eqalign{&Z=\int e^{-S_{CFT}}\biggl[  1-\mu^\epsilon\int g(F(x)\mu)\CO(x) d^4x +\half\mu^{2\epsilon}\int g(F(x)\mu)g(F(y)\mu)\CO(x)\CO(y) d^4xd^4y\cr&-{1\over 6}\mu^{3\epsilon}\int g(F(x)\mu) g(F(y)\mu) g(F(z)\mu)\CO(x)\CO(y)\CO(z) d^4xd^4yd^4z  +\cdots\biggr] ~.}}
Our goal is to find the dilaton dependence of the partition function. The terms in parentheses can be interpreted as 1-, 2-, and 3-point correlation functions of $\CO$s, evaluated in the conformal field theory. We can drop the term linear in $\CO$ because the expectation value of $\CO$ vanishes in the CFT. Thus we get
\eqn\CFTdili{\eqalign{&Z=\int e^{-S_{CFT}}\biggl[  1+\half\mu^{2\epsilon}\int g(F(x)\mu)g(F(y)\mu)\CO(x)\CO(y) d^4xd^4y\cr&-{1\over 6}\mu^{3\epsilon}\int g(F(x)\mu) g(F(y)\mu) g(F(z)\mu)\CO(x)\CO(y)\CO(z) d^4xd^4yd^4z+\cdots\biggr]~. }}
The remaining terms in parentheses are just the conventional 2- and 3-point functions in the CFT, so we get that 
 the partition function of the deformed theory differs from the partition function of the original CFT by
\eqn\CFTdilii{\eqalign{&\half\mu^{2\epsilon}\int {g(F(x)\mu)g(F(y)\mu)\over (x-y)^{8-2\epsilon}}d^4xd^4y-{C\over 6}\mu^{3\epsilon}\int {g(F(x)\mu) g(F(y)\mu) g(F(z)\mu)\over (x-y)^{4-\epsilon}(x-z)^{4-\epsilon}(y-z)^{4-\epsilon}}   d^4xd^4yd^4z+\cdots~. }}

In order not to jeopardize the expansion in $g$, we only integrate over domains where the distances between points are not parametrically different from $\mu$.  In this way, we will eventually complete integrating over the whole space by employing RG transformations. (Instead of  this conventional RG trick, one can also explicitly sum the leading contributions from {\it all} orders in perturbation theory.)

One can simplify~\CFTdilii\ drastically if one observes that expanding in $\tau$ is closely related to the expansion in $\epsilon$. This is due to  the following chain rule identity 
\eqn\tauder{{\del\over \del x} g(F\mu)=\beta(F\mu){\del\over \del x}\log F=\beta{\del\tau\over\del x}~.}
Notice that $g$ is of order $\epsilon$, while $\beta$ is of order $\epsilon^2$. In general, we pay a factor of $\epsilon$ for every additional $\tau$ that we extract. 

Let us now show how terms in the effective action for $\tau$ arise from the integrals in~\CFTdilii. We shall analyze the term $\mu^{2\epsilon}\int {g(F(x)\mu)g(F(y)\mu)\over (x-y)^{8-2\epsilon}}d^4xd^4y$ and later explain why the others are negligible to leading nontrivial order in the $\epsilon$ expansion. Expanding in derivatives we have  
\eqn\spaceexp{\eqalign{&\int {g(F(x)\mu)g(F(y)\mu)\over (x-y)^{8-2\epsilon}}d^4xd^4y\cr&=\int \left({\left(g(F(x)\mu)\right)^2\over(x-y)^{8-2\epsilon}}+{1\over8}{g(F(x)\mu)\square g(F(x)\mu)\over (x-y)^{6-2\epsilon}}+{1\over 192}{g(F(x)\mu)\square^2g(F(x)\mu) \over (x-y)^{4-2\epsilon}}+\cdots\right)d^4xd^4y
}}
Next, we integrate over $y$. We focus on the four-derivative term (namely the one appearing with coefficient $1/192$ in~\spaceexp). Performing the $y$ integral over an infinitesimal energy slice $(\mu+d\mu)^{-1}<|x-y|<\mu^{-1}$, plugging back to~\CFTdilii, and expanding in $\epsilon$  we obtain the following contribution to the effective action 
\eqn\collecting{ -{\Omega_3\over 384}\int d^4x
 g(F(x)\mu)\square^2g(F(x)\mu)d\log\mu~.}
Integrating by parts it becomes manifest that this contribution is of order $\epsilon^4$.  Indeed, focusing on the leading contribution in $\epsilon$,~\collecting\ is equivalent to
\eqn\effac{-{\Omega_3\over 384}
 \beta^2(g)d\log\mu\int d^4x (\square \tau)^2~.}

Using similar considerations one can verify  that the term proportional to $C$ in~\CFTdilii\ and all the higher corrections contribute only at a higher order in $\epsilon$. Hence,~\effac\ represents the genuine four-derivative effective action for the dilaton background field to leading order in $\epsilon$. 
We can now compare~\effac\ to the most general allowed four-derivative effective action for the dilaton~\geneffaction. Even though there are three equations in two variables, there is a solution with $\alpha_3=2(a_{UV}-a_{IR})$. Hence, the difference between the $a$-anomalies is  obtained by dividing the coefficient in~\effac\ by a factor 2 and integrating over all $\mu$
\eqn\changea{\Delta a=-{\Omega_3\over 768}\int \beta^2(g(\mu))d \log\mu~.}
 This can be equivalently written as an integral over the coupling $g$, ranging from the fixed point in the ultraviolet  with $g=0$ down to the CFT in the IR with $g_*=2\epsilon/C\Omega_3$
\eqn\changeai{\Delta a=-{\Omega _3\over 768}\int_{0}^{g_*} \beta(g) dg={1\over 1152}{\epsilon^3\over C^2\Omega_3}~.}

We have deduced the change in the $a$-anomaly from the Wess-Zumino term for the background coupling constant.  A more familiar definition of the $a$-anomaly is via the partition function over the four-sphere. In the next subsection we perform the path integral over the four-sphere explicitly and compare to~\changeai. 

\subsec{The Four-Sphere Partition Function}

Since the $a$-anomaly is proportional to the Euler density while the $c$-anomaly is proportional to the Weyl tensor squared, the $a$-anomaly can be isolated by computing the partition function over $S^4$.

As long as the perturbation is weak, one can compute the partition function by expanding in the coupling (in the spirit of~\CFTdil). This computation has been done in~\KlebanovGS\ and in a slightly different context originally in~\CardyCWA. Keeping only the significant terms in the weak coupling expansion,~\KlebanovGS\ found
\eqn\detlaF{\delta F(\lambda)=-{\lambda^2\over 2}{(2R)^{2\epsilon}\pi^{d+\half}\over 2^{d-1}}{\Gamma(-{d\over2}+\epsilon)\over \Gamma({d+1\over2 })\Gamma(\epsilon)}+{\lambda^3C\over 6}
{8\pi^{3(d+1)/2}R^{3\epsilon}\over \Gamma(d)}{\Gamma({-d\over 2} +{3\epsilon\over 2} )\over \Gamma({1+\epsilon\over 2})^3}~, }   
where $R$ is the radius of the $d$-dimensional sphere. To compute the $a$-anomaly we need to study the coefficient of $\log R$.\foot{Since there is no logarithmic dependence on the radius in odd-dimensional conformal field theories, the computation in~\KlebanovGS\ proceeds quite differently.} 
\eqn\detlaFi{{d\over d\log R}\delta F(\lambda)=-\lambda^2\epsilon{(2R)^{2\epsilon}\pi^{d+\half}\over 2^{d-1}}{\Gamma(-{d\over2}+\epsilon)\over \Gamma({d+1\over2 })\Gamma(\epsilon)}+{\lambda^3C\over 2}\epsilon
{8\pi^{3(d+1)/2}R^{3\epsilon}\over \Gamma(d)}{\Gamma({-d\over 2} +{3\epsilon\over 2} )\over \Gamma({1+\epsilon\over 2})^3}~. }   
If $d$ is even, the Gamma functions are singular around $\epsilon=0$, so one has to exercise a little care when expanding in $\epsilon$. The leading order result is
\eqn\detlaFi{{d\over d\log R}\delta F(\lambda)={(-)^{d/2}\over (d/2)!}\left(-\lambda^2\epsilon{\pi^{d+\half}\over 2^{d-1}\Gamma({d+1\over2 })}+{\lambda^3C\over 3}
{8\pi^{3d/2}\over \Gamma(d)} \right)~.}   

The next step is to express this result in terms of the physical coupling, which depends on the renormalization scale $\mu$. The relation between the bare and physical coupling is
\eqn\relcouplings{\lambda\mu^{-\epsilon} =g+{C\pi^{d/2} \over \epsilon \Gamma(d/2)}g^2+\CO(g^3)~.}
We plug this back into our partition function and keep only the leading terms.
We find 
\eqn\detlaFiii{\eqalign{&{d\over d\log R}\delta F(g)={(-)^{d/2} \pi^d\over 2^{d/2-2} (d/2)!(d-1)!! } \left(-{g^2\over2}\epsilon+{g^3C\Omega_{d-1}\over 6 }\right)  ~.}}
Note that the object in parenthesis is precisely the integral of the beta function, hence, the answer is stationary for fixed points, as it should be. 

The relation between the $\log R$ derivative of the partition function and the $a$-anomaly is through the Euler characteristic of the four-sphere, $\int_{S^4} E_4\sqrt g d^4x=64\pi^2$. This allows us to extract the formal quantity $\Delta a(g)$ 
\eqn\summfor{\Delta a(g)={ \pi^2\over 384 } \left({g^2\over2}\epsilon-{g^3C\Omega_{d-1}\over 6 }\right) ~,} 
which is generally an ambiguous object. However, when we set $g=g_*$ this describes the physical change in the $a$-anomaly as we flow from $g=0$ at short distances to the IR fixed point $g=g_*$. Substituting $g=g_*$ in~\summfor, one verifies that $\Delta a(g_*)$ coincides perfectly with the total change in the $a$-anomaly computed via the Wess-Zumino procedure in~\changeai.

\newsec{The Banks-Zaks Fixed Point}

In some respects, the Banks-Zaks fixed point~\BanksNN\ is quite similar to the weakly relevant flows analyzed in the previous section. However, it is sufficiently different to necessitate a separate treatment. 
The first marked difference is that in the Banks-Zaks fixed point one perturbs by a marginal (and {\it not} a slightly relevant) operator.  The weakly coupled fixed point is then achieved due to a balance between the one- and two-loop contributions to the beta function. The more conceptual difference is that the the Banks-Zaks fixed point is at infinite distance away from the free theory (distances are measured using the Zamolodchikov metric). Indeed, the perturbation of the free Yang-Mills theory by turning on a gauge coupling amounts to adding the operator $F_{\mu\nu}^2$ with a coefficient that scales like $1/g^2$. By contrast, the cases we studied in the previous section can be realized by adding to an existing CFT a well-defined (relevant) operator with an arbitrarily small coefficient. 

We define the gauge coupling $g$ in the usual way, through the Lagrangian $-{1\over 2 g^2 }\Tr(F_{\mu\nu}^2)$, where  $\Tr (T^AT^B)=\half \delta^{AB}$ in the fundamental representation. Then, in terms of $\alpha={g^2\over 16\pi^2}$ the beta function is 
\eqn\genbeta{{d\alpha\over d\log\mu}=\beta_0 \alpha^2+\beta_1\alpha^3+\cdots~,}
with the coefficients (for an $SU(N)$ gauge theory with $N_f$ Dirac fermions in the fundamental representation)
\eqn\betafull{-\half\beta_0={11\over3}N-{2\over 3}N_f~,\qquad  -\half\beta_1= {34\over 3}N^2-N_f\left({10\over 3} N+{N^2-1\over N} \right )~.}
In the large $N$ limit this simplifies to
\eqn\betalargeN{-\half\beta_0={11\over3}N-{2\over 3}N_f~,\qquad  -\half\beta_1= {34\over 3}N^2-{13\over 3}N_f N~. }
The Banks-Zaks limit is  $N,N_f\rightarrow\infty$, while $\epsilon={{11\over 2}N-N_f\over N}\ll 1$. In this case the beta function further reduces to 
\eqn\BZfp{{d\alpha\over d\log\mu}=-{4N\over3}\epsilon \alpha^2+25N^2\alpha^3+\cdots~.}
In the Banks-Zaks limit there is a controlled fixed point (with a parametrically small 't Hooft coupling) at $\alpha_*={4\epsilon\over 75 N}$. 

Along the flow one can choose the effective theory at a scale $\mu$ to take the form 
\eqn\lagi{\CL=-{1\over 2 g^2(\mu)} \Tr(F_{\mu\nu}^2)+\sum_{i=1}^{N_f}\bar\psi^i\gamma^\mu D_\mu\psi^i~. }
This form is achieved by appropriately rescaling the fermions. 
In this case $g$ represents both the self coupling of the gauge fields as well as the fermion-fermion-gauge vertex. If the beta function of the gauge coupling $g$ in~\lagi\ is zero then the theory is clearly conformal. This is why the beta function of $g$ chosen in this way coincides with the beta function~\BZfp\  to the order we are interested in. 

We can render this theory conformal by simply writing (as before)
\eqn\lagii{\CL=-{1\over 2 g^2(F\mu)} \Tr(F_{\mu\nu}^2)+\sum_{i=1}^{N_f}\bar\psi^i\gamma^\mu D_\mu\psi^i~,}
with $F=e^{\tau}$.
We can extract the dependence on $\tau$ by expanding the Lagrangian around $\tau=0$
\eqn\lagiii{\CL=-{1\over 2 g^2(\mu)} \Tr(F_{\mu\nu}^2)+\sum_{i=1}^{N_f}\bar\psi^i\gamma^\mu D_\mu\psi^i-\half\tau\beta_\lambda \Tr(F_{\mu\nu}^2)-{1\over4}\tau^2\dot\beta_\lambda\Tr(F_{\mu\nu}^2)+\cdots~.}
Here $\lambda=1/g^2$ and accordingly $\beta_\lambda\equiv {d\over d\log\mu}{1\over g^2}$. Expanding the functional integral in $\tau$,  we find that the leading nontrivial term is quadratic in $\tau$
\eqn\BZint{{1\over 8}\biggl\langle \int d^4xd^4y \ \tau(x)\beta_\lambda \Tr(F_{\mu\nu}^2(x))\tau(y)\beta_\lambda \Tr(F_{\mu\nu}^2(y))\biggr\rangle~.} 
Since the expansion in $\tau$ corresponds to an expansion in $\epsilon$ (as can already be seen in~\lagiii), it is  easy to see that all the corrections to~\BZint\ are suppressed by more powers of $\epsilon$. 

To obtain the answer to the leading nontrivial  order, we ought to evaluate the correlator $\bigl\langle  \Tr(F_{\mu\nu}^2(x))\Tr(F_{\mu\nu}^2(y))\bigr\rangle$ in free-field theory. The answer is 
\eqn\twopoint{\bigl\langle   \Tr(F_{\mu\nu}^2(x))\Tr(F_{\mu\nu}^2(y))\bigr\rangle ={12N^2g^4\over \pi^4(x-y)^8} ~.}
(This leads to a Zamolodchikov metric that scales like $ds^2\sim g^{-2}dg^2$. Hence, the distance to $g=0$ diverges logarithmically.)  Plugging this into~\BZint\ we find  
\eqn\BZint{{   3N^2\over 2\pi^4}g^4\beta_\lambda^2\int d^4xd^4y \ \tau(x) \tau(y)  {1\over (x-y)^8} ~.} 
Focusing on the four-derivative term (exactly as in the previous section) we get that the contribution to the effective action of the dilaton from the energy slice $d\log\mu$ is 
\eqn\BZinti{    { N^2\over 128\pi^4}\Omega_3g^4\beta_\lambda^2\int d^4x \ \tau(x)\square^2 \tau(x)  d\log\mu ~.} 

Since this is the complete four-derivative effective action to leading order in $\epsilon$, we can extract the change in the $a$-anomaly by comparing~\BZinti\ to~\geneffaction,\WZfourd.  Integrating over all energy scales one arrives at
\eqn\changeinaBZ{\Delta a={N^2\over 256\pi^4}\Omega_3\int_{\lambda_*}^{\infty}{d\lambda\over \lambda^2}\beta_\lambda~.}
Using \BZfp\ we can easily write the beta function for the coupling $\lambda$ 
\eqn\lambdabeta{\beta_\lambda= {4N\epsilon \over 48\pi^2}-{25\over (16\pi^2)^2}N^2\lambda^{-1}  ~.}
Note that the integral~\changeinaBZ\ converges at $\lambda=\infty$. The integral evaluates to 
\eqn\finalBZi{\Delta a={N^2\epsilon^2\over  3600 \pi^2}~.  }

The change in the $a$-anomaly of the Banks-Zaks fixed point has been computed in~\JackEB, directly from the definition of $a$.  To compare, we normalize~\finalBZi\ with respect to the  $a$-anomaly of a free scalar field ($a_{scalar}={1\over 90(8\pi)^2}$) and obtain 
\eqn\finalBZii{{\Delta a\over a_{scalar}}={8\over 5}N^2\epsilon^2~.  }
This precisely matches equation (5.19) of~\JackEB. 

The calculations in sections~2,3 can be generalized to other types of controllable fixed points. One could look at more general Banks-Zaks-like fixed points, for instance, ones that contain Yukawa couplings. Another interesting class of fixed points arises in supersymmetric models near points where the bottom component of some chiral operator has dimension $3-\epsilon$ (with $\epsilon>0$). Then there is a negative tree-level contribution to the beta function, but the 
``one-loop'' contribution is quite different from the ``one-loop'' contribution in models of the type described in section 2. In particular, in supersymmetric models (because of non-renormalization theorems) the ``one-loop'' contribution is cubic in the coupling (rather than quadratic as in~\betafun), and one can prove that it always has a positive sign~\GreenDA. Thus, if $\epsilon$  is small, a fixed point that can be studied in conformal perturbation theory always exists. This scenario is embedded naturally in many known supersymmetric models. One general way to realize it is to take any theory with a chiral operator whose dimension approaches $2$ from below and couple it to an external, free, singlet field.

\newsec{Comments on the Sum Rule~\sumrule}

Consider some conformal field theory describing the ultraviolet, and suppose this theory is deformed by relevant operators of dimensions $\Delta_i$.  The mass terms in front of these operators are promoted to functions of space time by $M\rightarrow Me^{-\tau (x)}$.  Therefore, the coupling of the spurion to the matter theory at high energies is always through terms with positive powers of various mass terms. 
Thus, dimensional analysis implies that $Im\CA(s)$ must behave, at large $s$, like \eqn\genrule{\lim_{s\rightarrow \infty}Im\CA(s)\sim s^{2-\epsilon}~,} where $\epsilon>0$. In terms of the dimension of the relevant operator in the UV, $\Delta_i$, $\epsilon=\min(4-\Delta_i)$. Hence, the sum rule is convergent in the UV.\foot{This argument leaves one possiblity unaccounted for -- when there is a marginally relevant operator. For instance, for a free Yang-Mills theory deformed by the gauge coupling one would get that the high energy behavior of the imagery part is $\sim s^2g^4(s)\sim s^2/\log^2(s) $. This is  sufficient for the convergence of the sum rule~\sumrule. In this section we will see that this is indeed the case for the Banks-Zaks flow.} 

Analogous arguments apply for the low $s$ properties of the imaginary part (one only needs to replace the relevant operators by irrelevant ones, because the IR CFT is approached by irrelevant operators which become less and less important at low energies). One therefore finds that the integral~\sumrule\ is IR convergent.\foot{Here the assumption that the IR theory is conformal (and not just scale invariant) appears vital. If the IR theory were only scale invariant, there would be a derivative coupling of the dilaton to the so called ``virial current,'' a non-conserved vector operator of dimension {\it three}.  By contrast, when the IR theory is conformal the dilaton couples only to irrelevant (or marginally irrelevant) operators in the CFT. Related ideas have been emphasized in~\NakayamaWQ.}

We will now study some examples of the imaginary part, $Im\CA(s)$. To warm up, consider the free massive scalar field. One replaces the mass by a space-time dependent function according to $M^2\rightarrow M^2e^{-2\tau}$. In~\KomargodskiVJ\ we have performed the path integral over the scalar field and showed that the generating functional for $\tau$ is of the form~\geneffaction,\WZfourd, with $a_{UV}-a_{IR}={1\over 90(8\pi)^2}$, the $a$-anomaly for a free scalar field. Let us now compute the imaginary part explicitly and examine the sum rule~\sumrule. Rewriting the term $\half M^2 e^{-2\tau}\Phi^2$ in terms of $\Psi$ we get
$-\half M^2\Phi^2+M^2\Psi\Phi^2-\half M^2\Psi^2\Phi^2$.

In the case of a free massive field, the diagrams contributing to the four-dilaton amplitude are all  one-loop diagrams. Their imaginary part is obtained by cutting them with the usual Cutkosky rules. In the forward configuration, the calculation is equivalent (by the optical theorem) to summing the  processes depicted in figure 4, squaring the amplitude, and integrating over the phase space. 

\medskip
\epsfxsize=3.5in \centerline{\epsfbox{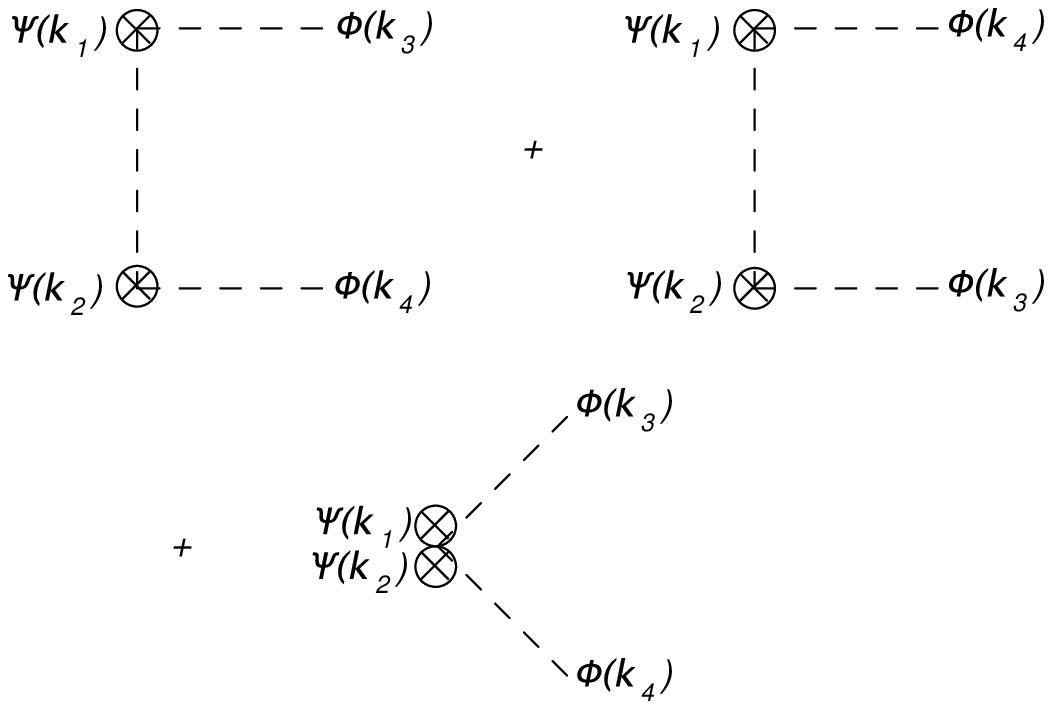}}
\noindent
Fig.4: Diagrams contributing to the imaginary part of the four-dilaton amplitude in the theory of a free massive scalar.
\medskip

The formula for the imaginary part is thus (one must not forget an extra factor of $\half$ in the optical theorem due to the fact we have identical particles in the final state)
\eqn\impart{Im\CA(s)={ |k_3|E_{c.m}\over 8|\epsilon_{\mu x y \nu}k_1^\mu k_2^\nu|   }{1\over (4\pi)^2  }\int d\Omega|\CM(k_1,k_2\rightarrow k_3,k_4)|^2~.}
(In the above we have assumed without loss of generality that the momenta of the $\Psi$s are along the $z$ direction.)  The imaginary part is manifestly positive, as required by unitarity. To compute the amplitude we sum over the diagrams in figure 4
\eqn\ampone{\CM=-2i M^2 - 4i{M^4\over (k_3-k_1)^2-M^2}-4i{M^4\over (k_4-k_1)^2-M^2}~.}
Denoting $k_1=(E,0,0,E)$, $k_2=(E,0,0,-E)$, we can parameterize the outgoing momenta as
$k_3=(E,k_\perp\cos\phi,k_\perp\sin\phi,|k|\cos\theta )$, $k_4=(E,-k_\perp\cos\phi,-k_\perp\sin\phi,-|k|\cos\theta)$, where $k_\perp=|k|\sin\theta$,
and the on-shell condition implies $E^2-M^2=|k|^2$. Thus the amplitude takes the form 
\eqn\amptwo{\half\CM=-i M^2 + {2iM^4E^2\over E^4\sin^2\theta+M^2E^2\cos^2\theta}~.}

The imaginary part as a function of the incoming particles' energy is thus 
\eqn\imparttwo{Im\CA(E)={  M^4\sqrt{E^2-M^2}\over  (4\pi)^2 E_{c.m}   }\int d\Omega  \left( 1-{2    \over {E^2\over M^2}\sin^2\theta+\cos^2\theta  } \right)^2~.}
Following the trivial integral over the azimuthal direction we integrate over $\theta$ and get 
\eqn\impartthree{\eqalign{&Im\CA(E)={   M^4\sqrt{E^2-M^2}\over  16\pi E  }\int_{-1}^1 d\cos\theta  \left( 1-{2    \over {E^2\over M^2}\sin^2\theta+\cos^2\theta  } \right)^2\cr& ={M^8 \over 8 \pi E^4}   \left(\left(2{E\over M^2} +{E^3\over M^4}\right)\sqrt{E^2-M^2}+\left(2-4{E^2\over M^2}\right)\tanh^{-1}\left({\sqrt{E^2-M^2} \over E }\right) \right)~.}}
Let us compare with our general comments earlier in this section~\genrule. Indeed, at  high energies~\impartthree\ behaves like $\CO(1)$, in perfect agreement with the anticipated result.

To extract the change in the $a$-anomaly we now integrate this over all energies via our sum rule~\sumrule\ 
\eqn\intsca{\Delta a={1\over 4\pi}\int_{s=4M^2}^\infty ds {Im \CA\over s^3}={1\over 32\pi}\int_{E=M}^\infty 
dE {Im\CA(E)\over E^5}~.}
This integral can be done easily and one finds $\Delta a={1\over 90(8\pi)^2}$. This is precisely the correct answer.

Let us now consider the imaginary part of the four dilaton amplitude in the  context of the Banks-Zaks flow. The vertices connecting the dilaton with the matter fields are given by the Lagrangian~\lagiii. First, we rewrite~\lagiii\  in terms of  $\Psi=1-e^{-\tau}$
\eqn\PsiBZ{\CL={-1\over 2g^2}Tr(F_{\mu\nu}^2)+\sum_{i=1}^{N_f}\bar\psi^i\gamma^\mu D_\mu\psi^i-\half\beta_\lambda
\left(\Psi+\Psi^2/2+\cdots \right)Tr(F_{\mu\nu}^2)-{1\over4}\dot\beta_\lambda(\Psi^2+\cdots) Tr(F_{\mu\nu}^2)~.}
To calculate the imaginary part of the four dilaton diagram we use again the optical theorem and relate $Im\CA$  to the transition of two insertions of $\Psi$ to anything.  Clearly, the leading contribution to such transitions is given by the vertex $-{1\over4}\beta_\lambda\Psi^2Tr(F^2_{\mu\nu})$ in~\PsiBZ. All the other vertices would contribute more factors of $\epsilon$. More explicitly, this vertex takes the form $ {1\over 8}\beta_\lambda\Psi^2(\del_\mu A^a_\nu-\del_\nu A^a_\mu)(\del_\mu A^a_\nu-\del_\nu A^a_\mu)$, where we neglected contributions from the non-Abelian terms since they entail more powers of $\epsilon$. See figure 5.

The imaginary part is thus given by (where the quantum numbers in the final state are $(a,\epsilon^{(1)})$, $(b,\epsilon^{(2)})$, and one also keeps in mind that the particles in the final state are identical) 
\eqn\ImBZ{Im\CA(E)={1\over 32(2\pi)^2}\int d\Omega \sum_{final \ states}\left|\CM(k_1,k_2\rightarrow (k_3,a,\epsilon^{(1)})(k_4,b,\epsilon^{(2)}) ) \right|^2~.}
After a short computation one finds that the amplitude is given by \eqn\ampBZ{\CM(k_1,k_2\rightarrow (k_3,a,\epsilon^{(1)})(k_4,b,\epsilon^{(2)}))=g^2 \beta_\lambda\left((k_3\cdot k_4)(\epsilon^{(1)}\cdot \epsilon^{(2)})-(k_3\cdot \epsilon^{(2)})(k_4\cdot \epsilon^{(1)})    \right) \delta^{ab}~.}
Note that the beta function depends on the energy via the dependence of the gauge coupling on the scale $\mu$. In this case, again, the renormalization group allows us to write the answer without summing explicitly infinitely many diagrams. (Of course, this is precisely what the renormalization group is made for.)
A simple consistency check can be performed right away; we can verify that upon replacing the polarization vectors by momentum the amplitude vanishes. 

\medskip
\epsfxsize=1.8in \centerline{\epsfbox{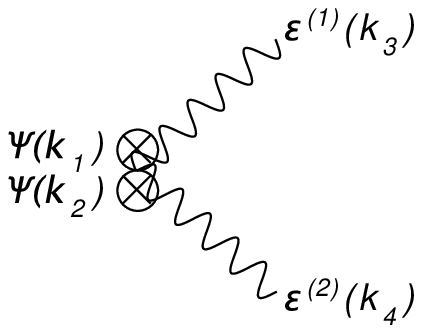}}
\noindent
Fig.5: Leading diagram contributing to the imaginary part of the four-dilaton amplitude in the Banks-Zaks flow.
\medskip

We now square the amplitude and sum over all the possible final states, that is, all the possible polarizations and color states. We find 
\eqn\sumfinal{\sum_{final \ states}|\CM|^2= 2g^4N^2\beta_\lambda^2(k_3\cdot k_4)^2~.}
We conclude that the imaginary part is 
\eqn\ImBZ{Im\CA(E)={N^2\over 4\pi} g^4(E) \beta_\lambda^2(E)E^4 ~.}
We extract the change in the $a$-anomaly very much like in~\intsca\ 
\eqn\changeBZ{\Delta a={N^2\over 128\pi^2}\int_0^\infty {dE\over E}g^4(E)  \beta_\lambda^2(E)={N^2\over 128\pi^2}\int_{\lambda_*}^\infty {d\lambda\over \lambda^{2}} \beta_\lambda~.}
Comparing to~\changeinaBZ\ we find perfect agreement. In addition, the high energy behavior of the imaginary part is consistent with our comments in the beginning of this section.

\goodbreak
\bigskip
\bigskip
\centerline{\bf Acknowledgments }

We would like to thank O.~Aharony, M.~Buican, T.~Dumitrescu, H.~Elvang, D.~Freedman, J.~Hung, D.~Jafferis, M.~Kiermaier, I.~Klebanov, J.~Maldacena, R.~Myers, S.~Pufu, B.~Safdi,  N.~Seiberg, J.~Sonnenschein, S.~Yankielowicz, and S.~Zamolodchikov  for many illuminating discussions directly and indirectly pertaining to this project. We are espeically indebted to A.~Schwimmer for asking many incisive questions throughout this work. In addition, we are grateful to S.~Theisen for providing us with very helpful unpublished notes.
Z.K. is supported by NSF PHY-0969448,  by a research grant from Peter and Patricia Gruber Awards, and by the Israel Science Foundation (grant $\#$884/11).

\listrefs
\end